# TapDrag: An Alternative Dragging Technique on Medium-Sized Multi-Touch Displays Reducing Skin Irritation and Arm Fatigue

Lasse Farnung Laursen
Independent Researcher
lasse@laursen.com

Hsiang-Ting Chen
University of Technology Sydney†
tim.chen@uts.edu.au

Paulo Silva
The University of Tokyo‡
paulo.fernando.silva@gmail.com

Lintalo Suehiro
Independent Researcher
lintalo@suehiro.dk

Takeo Igarashi
The University of Tokyo‡
takeo@acm.org

† University of Technology Sydney, City campus, 15 Broadway, Ultimo NSW 2007, Sydney, Australia
‡ The University of Tokyo, 7-3-1 Hongo, Bunkyo-ku, Tokyo, 113-0033, Tokyo, Japan

**ABSTRACT**

Medium-sized touch displays, sized 30 to 50 inches, are becoming more affordable and more widely available. Prolonged use of such displays can result in arm fatigue or skin irritation, especially when multiple long distance drags are involved. To address this issue, we present TapDrag, an alternative dragging technique that complements traditional dragging with a simple tapping gesture on both ends of the intended dragging path. Our experimental evaluation suggests that TapDrag is a viable alternative to traditional dragging with faster task completion times for long distances. Qualitative user feedback indicates that TapDrag helps prevent skin irritation. A reduction in arm fatigue remains unconfirmed.

**Keywords**: Drag, 2D target acquisition, large display, empirical evaluation, interaction design

**Index Terms**: H.5.2 [Information Interfaces and Presentation]: User Interfaces—Graphical user interfaces (GUI); H.5.2 [Information Interfaces and Presentation]: User Interfaces—Input devices and strategies, prototyping;

## 1 INTRODUCTION

Medium-sized touch-displays are becoming cheaper and more popular [1]. We define the medium-sized touch-display as having a diagonal screen length of approximately 30 to 50 inches, i.e. a screen within full reach of both arms of an average adult. These touch-devices are becoming more common in shared work environments to aid in design tasks [2], used as live performance instruments [3, 4], and used as public interactive installations [5, 6]. In these applications, the user will often be required to perform numerous dragging interactions, e.g. arranging photos/cards/media [2, 5, 6], and continuously positioning/adjusting audio/video elements [3, 4].

While developing prototype software on a medium-sized touch display, we observed that prolonged use could lead to both arm fatigue and skin irritation, especially when lengthy drag gestures were involved. Long distance dragging can be tiring because it involves extensive arm movement across the entire display, and the continuous friction between the finger and the display over an extended period of time is more likely to result in skin irritation. Novice users are more prone to experience skin irritation, as they may apply more pressure than required when interacting with an unfamiliar large touch display. As support for pressure sensitive interaction in popular touch devices continues to grow, e.g., force touch displays in recent Apple devices, these problems may be further compounded.

Previous works have addressed interaction issues on large displays using interface modifications [7, 8], novel widgets [9] and additional hardware [10, 11]. However, an alternative dragging technique that more closely mimics the simplicity of the traditional drag while also reducing skin irritation and arm fatigue, for medium-sized displays, could lead to a better interaction experience.

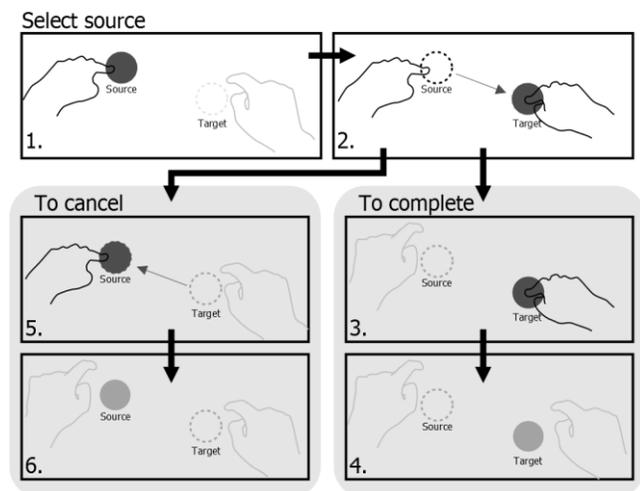

Figure 1: TapDrag interaction sequences.

We describe TapDrag, an alternative dragging technique that completes dragging tasks with simple tapping gestures on both ends. Traditional dragging gestures requires the user to place a finger at a source position, then slide it to the target position without leaving the display surface. TapDrag, as shown in Figure 1, achieves the same result by having the user place one finger at the source position, then another at the target position. The object then instantly moves from one finger to the other. To complete the TapDrag, the user lifts the finger at the source, and finally the finger at the target. Less sliding contact between the user's fingers and the touch surface mitigates skin irritation. Enabling users to use both hands to complete the interaction reduces the likelihood of full arm movement over a longer distance.

We conducted an experimental study, comparing TapDrag to traditional dragging. Overall, the results suggest that TapDrag is a good alternative to the traditional drag gesture for long distance dragging on medium-sized multi-touch displays. We implemented TapDrag alongside popular multi-touch gestures and found no conflicts in a single user scenario.

## 2 RELATED WORK

A growing body of work exists regarding interactive touch surfaces. Buxton et al. [12] presents one of the earliest related works, looking closer at issues and techniques for touch-interaction based devices. Buxton et al. explicitly note upon the issue of friction caused by using touch surfaces, but do not further explore the issue.

As larger displays have become more affordable over time, more research has focused on making interactions with such larger surfaces more effective. Czerwinski [13] provides an overview of many existing works, and we note some of the most closely related here.

A number of interaction techniques introduced in these works require additional hardware to function: Pick-and-Drop [10] moves GUI elements between multiple displays via stylus-based tapping gestures. Sugiura et al. extend this technique to function with a user's fingertips, using a finger print scanner [11].

Other techniques alter or introduce new interface elements: Baudisch et al. present Drag-and-Pop [7], an interaction technique usable for both pen and touch, where the user interface is altered to bring interact-able icons closer to a different icon the user is actively dragging. Collomb et al. improve upon previously presented techniques (Drag-and-Pop among others) with Push-and-Pop [8] designed specifically for wall-sized displays. Similar to the interaction techniques it is based on, Push-and-Pop also alters the user interface during usage. Kahn et al. present interaction techniques using a widget dubbed the `Frisbee' [9], designed for improved interaction on large displays. The `Frisbee' is very functional, but consequently also complex as an alternative for frequent dragging on a medium sized display.

Hoggan et al [14, 15] examine the performance and ergonomics of two common spatially fixed gesture types: Pinch and Rotation. They suggest optimal gesture settings, as opposed to introducing novel interaction techniques.

Wilson and Benko mention `Through-Body' transitions [16] which uses camera hardware, and Wilairat describes 'Drop target gestures' [17], both of which are similar to TapDrag. However, both descriptions are limited to an instance of moving a single target. The descriptions do not detail cancellation, multi-target movement, or integration with existing traditional gestures. Wilson and Benko informally evaluate `Through-Body' transitions, and Wilarat does not evaluate 'Drop target gestures'.

## 3 TAPDRAG

Our proposed technique - TapDrag - is designed for medium-sized multi-touch devices. Figure 1 illustrates two common interaction patterns using TapDrag to move an item. The interaction proceeds as follows: The user touches the source (item) with one finger (1), and then proceeds to touch the intended target with another finger (2), while the first finger remains on the source. Visually, the object will now instantly move from the first, to the second touch as visual feedback. The TapDrag is then completed by releasing the touches in the same order (3,4). To cancel the TapDrag the user instead releases the finger on the target first (5), and the source second. This is the reverse interaction required to complete the TapDrag. As users release their finger on the target, the object will be instantly moved back to indicate that the user is currently cancelling the TapDrag. The user can either then touch another target using the second finger, or fully abort from the operation by releasing the first finger (6). Figure 2 shows the complete state transition of the TapDrag interaction. Since TapDrag relies solely on tapping as its interaction method, it is less susceptible to detection issues caused by traditional dragging.

Our integration of TapDrag with pre-existing multi-touch gestures involves distinguishing between draggable objects and the background. The specifics of integrating TapDrag with other multi-touch gestures are detailed at the end of the paper.

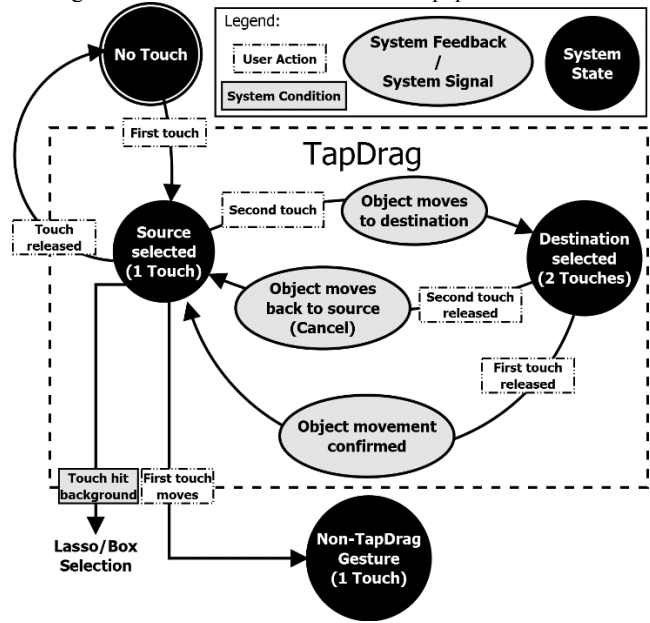

Figure 2: TapDrag state machine.

TapDrag can be used to examine multiple potential target destinations in a row without commitment by touching these destinations one by one with the second finger holding the first finger down. This is useful for adjusting a time slider in a video player. The user can also achieve lasso selection, as shown in Figure 3. The user touches the starting point with the first finger (1) and touches consecutive points along the intended lasso stroke by the second finger (1,2), and finally releasing the first finger after completing the lasso (3).

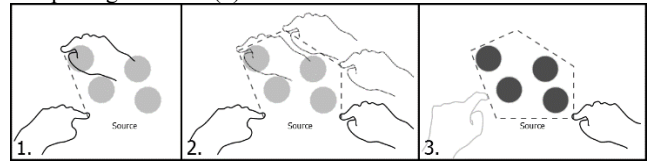

Figure 3: TapDrag lasso selection.

TapDrag also supports moving multiple items simultaneously. Using traditional dragging, several items are simultaneously moved by two sequential drags. The first drag operation is for box selection and second drag is for dragging of the selected objects. TapDrag also achieves this via two sequential TapDrags (total 4 touches), as shown in Figure 4. The first TapDrag completes the box selection (1). The second TapDrag is used to drag all of the selected objects to a new target (2).

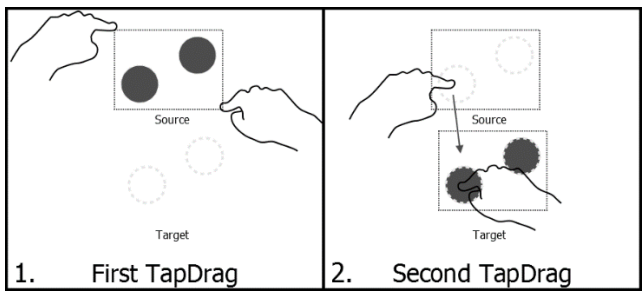

Figure 4: TapDrag box selection and movement.

## 4 EVALUATION

We conducted a user study comparing the performance of the basic TapDrag with traditional dragging in a series of single object dragging tasks.

### 4.1 Participants

A total of 18 participants (15 male and 3 female) between the ages of 21 and 45 were recruited. 17 were predominantly right-handed, and one was left-handed. All participants reported using touch-devices on daily basis.

### 4.2 Task and Conditions

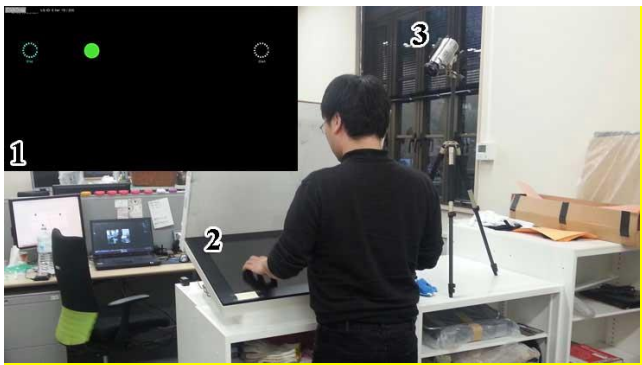

Figure 5: An image of the user study setup. 1. A screen-capture of a single long distance dragging task. 2. The touchscreen used throughout the user study. 3. The capture equipment used to record all of the users interactions.

To compare the two methods, each participant was asked to complete a series of dragging trials. All trials were performed on a Multitouch 3200 [18], a 32 inch multi-touch device, tilted towards the user, as shown in Figure 5. Schulz et al. found no overall optimal tilt angle [19], but their results indicate that a tilt towards the user is preferable.

In each trial, participants dragged an object (a circle with a diameter of ~3.5 cm) from a source position to a target position, as shown in the top left corner of Figure 5. Prior to commencing the actual study, users were allotted a few minutes to practice, so they felt comfortable using the dragging gestures to be tested. Prior to using TapDrag, each user was explained how to perform both a two handed and a one handed TapDrag. A complete study session lasted between 15-20 minutes, within the interaction times measured by Seto et al. in public spaces [6].

The study used a 2x2x2x2x4 with-in subject repeated measurement design with five factors: dragging type (TapDrag or traditional drag), target visibility (showing or not showing target area prior to first touch), source area (left-half or right-half), dragging distance (10cm short dragging or 55cm long dragging)
and dragging direction (up, down, left, right). Note that due to the limitations of the touch surface display size, long drags could only be tested from left to right, or right to left. Ten trials were recorded for each unique combination resulting in 7200 data points.

The short distance dragging trials were set to 10cm to ensure that the average adult could perform the TapDrag using only a single hand. The long distance dragging trials were set to 55cm, as this was the maximum distance the touch device could accommodate while still maintaining a generous border to randomize positioning. Note that the source area factor indicates the starting region lying on left half or right half of the display whereas the exact starting position are randomized for each factor combination.

These factors were chosen to cover common dragging patterns and to evaluate potential interesting usage patterns for TapDrag. For example, when designing TapDrag, we observed "arm crossing behavior", i.e. right hand touches source on left hand side, while left hand touches the target on the right, potentially increasing interaction difficulty. The target visibility factor helps test how often such behavior may occur. The source area factor encourages users performing the trial to occasionally use their non-dominant hand. All user study sessions were recorded with a video recorder and annotated manually to identify these special usage patterns.

## 5 RESULT

We analyse the task completion time with repeated measure ANOVAs and the failure rate with Friedman test. Completion times and failure rates are shown in Figure 6.

The average trial completion time for traditional drags and TapDrag were 1.55s and 1.49s, respectively. TapDrag was faster but there was no significant difference (F(1, 17)=1.064, ns). Unsurprisingly, both the target visibility and dragging direction factor had main effects ($p<0.01$).

The dragging type factor had an interaction with dragging distance. For short dragging, the average completion time for traditional drags and TapDrag were 1.44s and 1.47s, respectively, with no significant difference (F1, 17=1.76, ns). For longer dragging, the completion time were 2.01s and 1.59s, respectively, with TapDrag being significantly faster (F1, 17=31.975, $p<0.01$). The failure rates for traditional drags and TapDrag were 0.03 and 0.08 respectively, and with a significant effect ($\chi2(1,N=18) = 30.38$, $p < .001$). Looking closer, the dragging type only had significant effect for short dragging ($\chi2(1,N=18) = 18$, $p < .001$), but not for long dragging ($\chi2(1,N=18) = .286$, ns).

When asked about the preference of dragging technique in the interview session, 11 out of the 18 user study preferred TapDrag overall, although this might be skewed by the novelty factor.

### 5.1 Discussion and observations

Overall the results indicated that TapDrag could be a useful technique for long distance dragging, where as traditional drags are more suited for shorter distances. A number of participants noted on their own accord that a combination of the traditional drag and TapDrag would be preferable.

The Multitouch 3200 hardware specifications denote a response rate of 100 Hz when when less than 20 simultaneous touches are occurring. However, we observed a higher latency despite receiving and interpreting input directly from the operating system driver layer. Our observations indicate that latency remains an issue with larger display devices, and it is important to acknowledge that this delay may introduce a positive bias in favour of TapDrag.

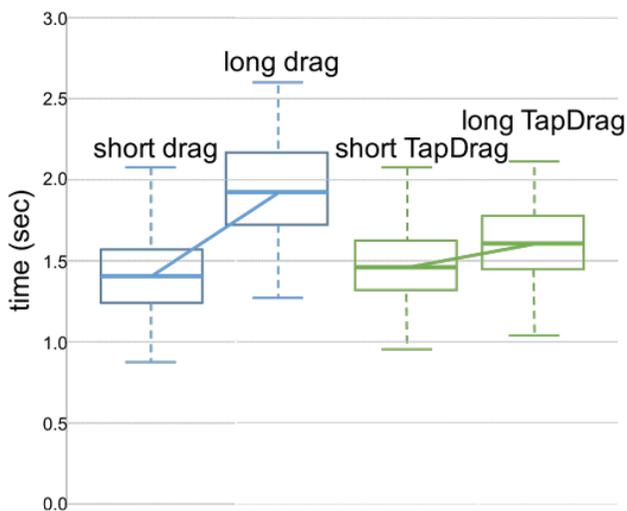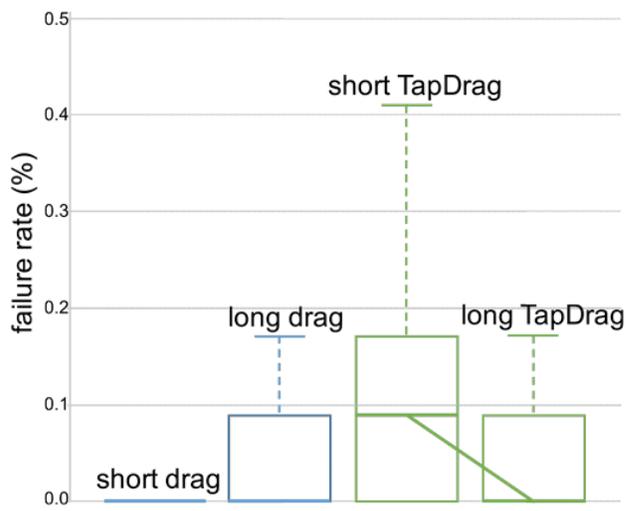

Figure 6: A Tukey box plot of trial completion times and failure rates. Whiskers designate the lowest and highest measured value.

### 5.1.1 User Discomfort

While performing tasks using traditional dragging, 10 participants reported experiencing skin irritation, and two participants reported experiencing arm fatigue, out of 18 participants. No participants reported experiencing skin irritation while using TapDrag, and one participant reported experiencing arm fatigue.

The overall low reports of arm fatigue indicates that this issue is only of concern for longer periods of use with larger touch devices.

Users most often reported skin irritation when performing an upward traditional drag. We observed this occurring when the users performed this upward drag using their index finger pointed in the same direction as the movement. Several factors contribute to this irritation: tilt-angle of the touch device, pressure applied by the users, and the lack of excess skin to stretch (being affixed to the nail). Users would often deliberately twist their hands to align it sideways with the touch surface, when performing a traditional upwards drag. By doing this, users could effectively drag objects upwards while moving their finger sideways. TapDrag avoids this uncomfortable interaction altogether by not requiring the user to actually drag their finger along the touch surface.

Multiple users reported experiencing their finger skipping across the surface of the display, during long traditional drag tasks. We experienced that long traditional drags erode the natural skin moisture, causing increased friction and possibly further discomfort. By design, TapDrag minimizes skin/surface contact, which consequently minimizes erosion to the user's natural skin moisture preventing skin irritation.

### 5.1.2 TapDrag Usability

When performing tasks using traditional dragging, we observed 10 participants using their dominant hand exclusively, and the remaining 8 switching in-between both. TapDrag requires bimanual input by design. For long distances, it forces the user to distribute the workload between both hands. Only one participant noted a perceived overall increase in cognitive load when using TapDrag. This requires further investigation.

Using two hands increases the chance that the users may occlude relevant interface elements during use. We observed this occurrence rarely and only in instances where the target area was not initially visible. Our observations indicate that this is a non-issue, with interfaces users are familiar with.

### 5.1.3 Failed drags

The measured failure rate for short distance TapDrags is more than three times that of long distance TapDrags. Previous research by Forlines et al. [20] reported a higher selection error correlating with longer distances between targets, during bimanual touch operation. This stands in opposition to our own findings. This is partially due to the fact, that Forlines et al. measure error percentages, while we use a simpler binary pass/fail measurement with a significant tolerance (~3.5 cm diameter). Furthermore, Forlines et al. require the user to operate two inputs simultaneously, while TapDrag's interaction pattern is strictly linear.

We observed that the majority of failed short distance TapDrag's occurred during single hand use. The user would sometimes erroneously lift the source touch prior to the touch-surface registering the target touch. Although multiple users noted that, they expected to perform better with more practice, TapDrag appears most beneficial for long distance drags, offering comparable error rates and shorter completion times.

## 6 INTEGRATING TAPDRAG INTO MULTI-TOUCH SYSTEM

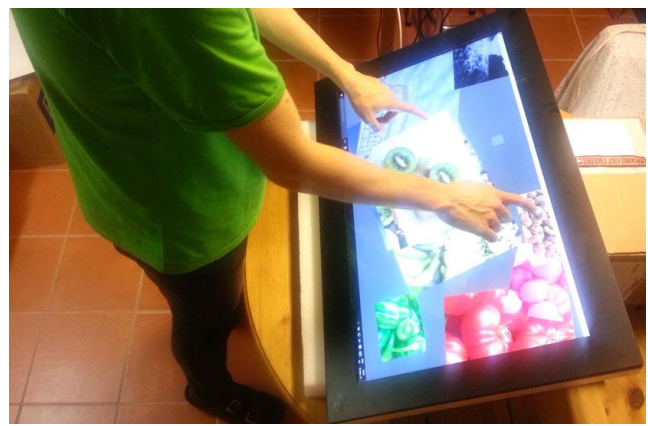

Figure 7: TapDrag and other multi-touch gestures used in a photo arranging application.

TapDrag is compatible with common multi-touch gestures, such as standard dragging, pinching, spreading, and rotation. We implemented all of the aforementioned gestures alongside TapDrag in a photo arrangement application, shown in Figure 7, and featured in the supplemental video. In the application, a user can freely arrange pictures with both traditional dragging and TapDrag gestures and control image size and orientation with common two-fingers pinching and rotation gestures.

To solve potential conflicts between TapDrag and other multi-touch gestures, a simple solution is to assume that if both touches hit upon the same object, then the user is attempting to use traditional gestures, such as zooming, rotating, and moving. Currently our gesture recognition engine adopts this solution. Figure 8 shows the complete state diagram for this approach. Note that this solution implies short distance TapDrags inside the boundary of an object are automatically rejected. Since we encourage users to use traditional dragging for short distance movement, this is a desirable compromise.

It is possible to distinguish between rotation/zoom/move and TapDrag, even if one of the touches occur outside the boundary of an object. If either of the two touches move, it cannot be a TapDrag gesture. However, this approach has an ambiguous state when the second touch occurs: at this point, the system should - in case of a TapDrag - preview the intended destination (see Figure 1, step 2). But prior to letting go of the inital touch, or moving either of the two touches, both TapDrag and rotation/zoom/move gestures are possible. In this case, a possible compromise would be to use a ghost/transparrent preview. If the user then proceeds to move either of the touches a zoom/rotate/move gesture is triggered, and the TapDrag can be cancelled.

A remaining ambiguity is when the users place their first touch on a draggable object (without moving it), and the second touch on a different draggable object. This scenario can be interpreted as either a single TapDrag, or two separate traditional drags, which have yet to commence. Our observations indicate that users rarely perform bimanual interaction with two separate objects, which can be manipulated via a single hand. Therefore, unless the system should not support placing one item on top of another, we would recommend triggering a TapDrag in a single user scenario.

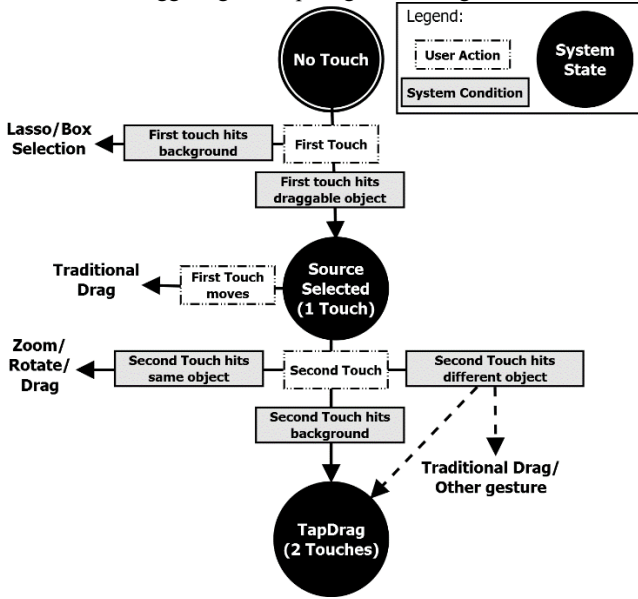

Figure 8: Integrating basic TapDrag with pre-existing multi-touch gestures.

Similar to standard pinch gestures, TapDrag is not easily distinguished in a multi user scenario. Without being able to determine which touches belong to which user, large distances between touches will lead to significant detection ambiguity.

## 7 CONCLUSION

The increasing availability of bigger sized touch displays warranted a closer look at alternative touch interaction methods suited for dragging, which minimize skin irritation and arm fatigue. TapDrag can integrate with existing popular multi-touch gestures, as well as the traditional drag gesture. The results provided evidence that TapDrag can be a suitable replacement for traditional dragging when long distance drags are required. Consequently, we suggest integrating TapDrag along-side the traditional drag when long distance drags are required.

### 7.1 Future work

TapDrag opens up a number of interesting directions of further research.

Similar to the popular pinch gesture, TapDrag can involve two tap gestures with a significant distance in-between. In a multi-user scenario, this can lead to erroneous behavior if a tap is attributed to the wrong user. Ideal future work would involve finding a solution for error-free multi-user TapDrag which does not require additional hardware, as such requirements can be unsuitable a public setting. One possibility might involve dragging the initial tap gesture a little towards the intended target displaying a narrow cone within which only a secondary tap would be detected.

Pressure sensitive touch devices are emerging which may compound existing issues of user discomfort. For touch devices supporting pressure detection TapDrag could be an interesting avenue of investigation.

Applying TapDrag to emerging augmented reality interfaces such as Microsoft's HoloLens would be another promising body of work. These types of interfaces can involve multiple successive full arm movements and may not even have a traditional surface with which to drag elements across.

Uncommon gestures are occasionally implemented in specialized touch controlled software. Integrating TapDrag with these less widespread gestures warrants further investigation.

TapDrag has a low discoverability compared to traditional dragging which is a very intuitive gesture. The optimal way to communicate TapDrag to users of varying skill is an ideal avenue of future work.

## 8 ACKNOWLEDGEMENTS

We would like to extend our thanks to all of the participants who helped us conduct the user study. This work was supported by JSPS KAKENHI Grant Number 24-02734, and a generous donation from MSRA. A significant amount of this work was conducted on premises and in collaboration with the University of Tokyo.